\documentclass[11pt]{article}
\usepackage{emnlp2016}
\usepackage{times}
\usepackage{latexsym}
\usepackage{amssymb}
\usepackage{amsmath}
\usepackage{graphicx}
\usepackage{multirow}
\usepackage{textcomp}
\usepackage{url}
\usepackage{array}
\usepackage[caption=false]{subfig}
\usepackage{color,comment}

\emnlpfinalcopy 

\title{Learning Latent Local Conversation Modes for Predicting Community Endorsement in
Online Discussions}

\author{Hao Fang \quad Hao Cheng \quad Mari Ostendorf \\
	University of Washington \\
	\tt{\{hfang,chenghao,ostendorf\}@uw.edu}}

\date{}

\begin{document}

\maketitle

\begin{abstract}
	Many social media platforms offer a mechanism for readers to react to
	comments, both positively and negatively, which in aggregate can be thought of
	as community endorsement.
	This paper addresses the problem of predicting community endorsement in online
	discussions, leveraging both the participant response structure and
	the text of the comment.
	The different types of features are integrated in a neural network that
        uses a novel architecture to learn latent modes
	of discussion structure that perform as well as deep neural networks but are
	more interpretable.
	In addition, the latent modes can be used to weight text features thereby
	improving prediction accuracy.
\end{abstract}

\section{Introduction}
Online discussion forums provide a platform for people with shared interests
(online communities) to discuss current events and common concerns. Many
forums provide a mechanism for readers to indicate positive/negative
reactions to comments in the discussion, with up/down votes, ``liking,'' or
indicating whether a comment is useful. The cumulative reaction, which we
will refer to as ``community endorsement,'' can be useful to readers for
prioritizing what they read or in gathering information for decision making.
This paper introduces the task of automatically predicting the level of
endorsement of a comment based on the response structure of
the discussion and the text of the comment. To address this task, we introduce a
neural network architecture that learns latent discussion structure (or,
conversation) modes and adjusts the relative dependence on text vs.\ structural
cues in classification. The neural network framework is also useful for
combining text with the disparate features that characterize the submission context of a comment, i.e.\ relative timing in the discussion, response structure (characterized by graph features), and author indexing. 

The idea of conversation modes stems from the observation that regions of a discussion can be qualitatively different: low vs.\ high activity, many participants vs.\ a few, etc. Points of high activity in the discussion (comments that elicit many responses) tend to have higher community endorsement, but some points of high activity are due to controversy. We hypothesize that these cases can be distinguished by the submission context, which we characterize with a vector of graph and timing features extracted from the local subgraph of a comment. The context vectors are modeled as a weighted combination of latent basis vectors corresponding to the different modes, where bases are learned using the weak supervision signal of community endorsement. We further hypothesize that the nature of the submission context impacts the relative importance of the actual text in a comment; hence, a mode-dependent gating mechanism is introduced to weight the contribution of text features in estimating community endorsement.


The model is assessed in experiments on Reddit discussion forum data, using
karma (the difference in numbers of up and down votes) as a proxy for community
endorsement, showing benefits from both the latent modes and the gating. As
described further below, the prediction task differs somewhat from prior work on
popularity prediction in two respects. First, the data is not constrained to
control for either submission context or comment/post content, but rather the
goal is to learn different context modes that impact the importance of the
message. Second, the use of the full discussion thread vs.\ a limited time
window puts a focus on participant interaction in understanding community
endorsement.

\section{Related Work}
The cumulative response of readers to social media and online content has been studied using a variety of measurements, including: the volume of comments in response to blog posts \cite{Yano2010ICWSM} and news articles \cite{Tsagkias+09,Tatar+11}, the number of Twitter shares of news articles \cite{Bandari+12}, the number of reshares on Facebook \cite{Cheng2014WWW} and retweets on Twitter \cite{Suh+10,Hong+11,Tan2014ACL,Zhao2015KDD}, and the difference in the number of reader up and down votes on posts and comments in Reddit discussion forums \cite{Lakkaraju2013ICWSM,Jaech2015EMNLP}. An advantage of working with the Reddit data is that both positive and negative reactions are accounted for, so the total (karma in Reddit) is a reasonable proxy for community endorsement.

For all the different types of measures, a challenge in predicting the
cumulative reaction is that the cases of most interest are at the tails of a
Zipfian distribution. Various prediction tasks have been proposed with this in
mind, including regression on a log score \cite{Bandari+12}, classification into
3-4 groups (e.g. none, low, high) \cite{Tsagkias+09,Hong+11,Yano2010ICWSM}, a
binary decision as to whether the score will double given a current score
\cite{Lakkaraju2013ICWSM}, and relative ranking of comments
\cite{Tan2014ACL,Jaech2015EMNLP}. In our work, we take the approach of
classification, but use a finer grain quantization with bins automatically
determined by the score distribution.

The work on cumulative reaction has mostly considered two different scenarios: predicting responses before a comment/document has been published vs.\ after a limited lookahead time for extracting features based on the initial response. While the framework proposed here could handle either scenario, the experiments reported allow the classifier to use a longer future window, until most of the discussion has played out. This provides insight into the difficulty of the task and illustrates that volume of responses alone does not reliably predict endorsement.

A few studies investigate language factors that may impact popularity through carefully controlled experiments.
To tease apart the factor of content quality, \newcite{Lakkaraju2013ICWSM} predict resharing of 
duplicated image submissions, investigating both the submission context (community, time of day, resubmission statistics) and
language factors. Our work differs in that content is not controlled and the submission context includes the response structure and relative timing of the comment within the discussion.
%
\newcite{Tan2014ACL} futher control
the author and temporal factors in addition to the topic of the content,
by ranking pairs of tweets with almost identical content made by the same
author within a limited time window.
\newcite{Jaech2015EMNLP} control the temporal factor for ranking Reddit comments
made in a time-limited window and study different language factors. 
Here, rather than manually controlling the submission context, we
propose a model to discover latent modes of submission context (relative timing, response structure) and analyze its utility in predicting community endorsement.
Furthermore, we study how the usefulness of language information in
estimating the community endorsement varies depending on submission context.

\section{Data and Task}
\label{sec:data_task}
\begin{figure}[t]
	\centering
	\includegraphics[width=0.33\textwidth]{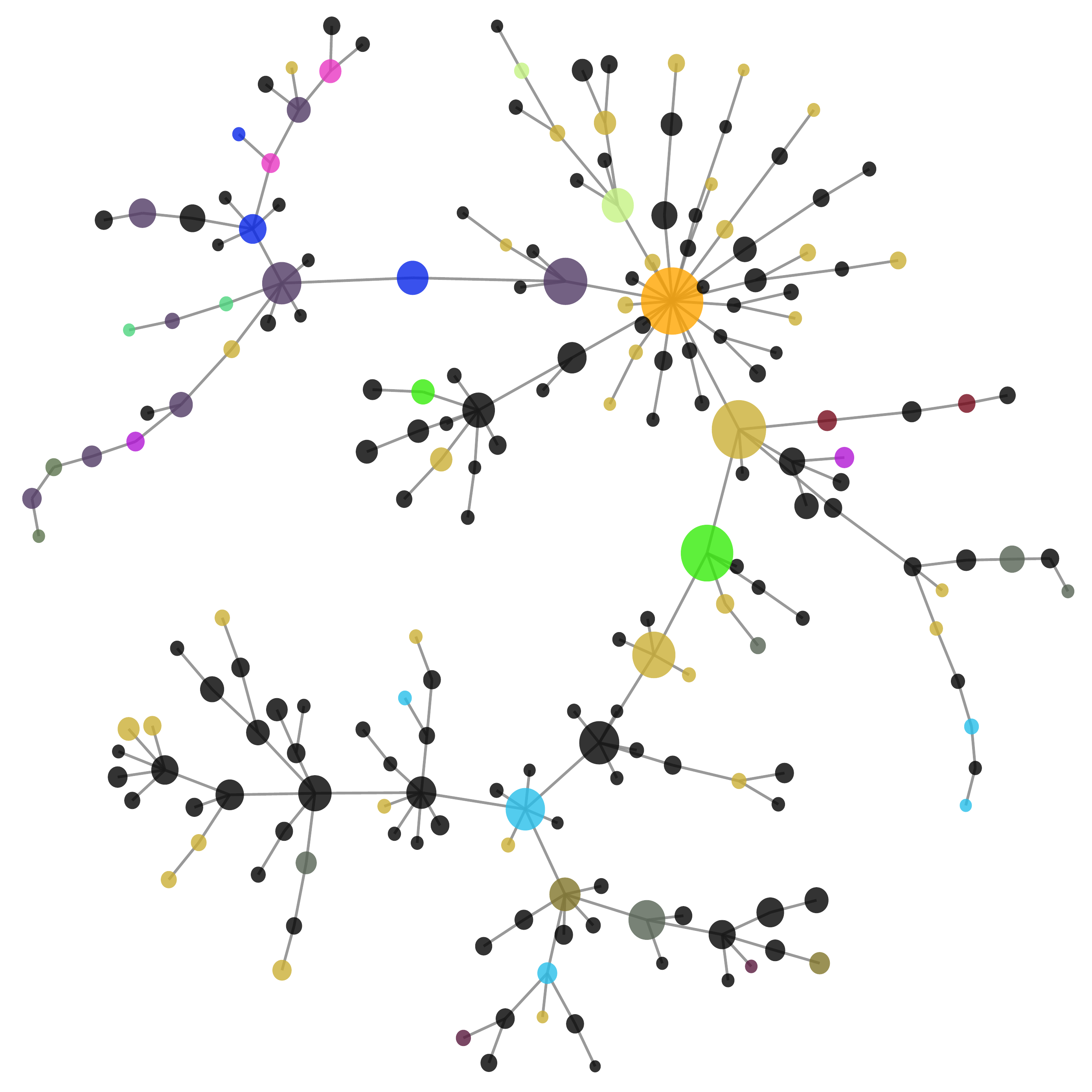}
	\caption{Visualization of a Reddit discussion thread.
		The orange node represents the root post;
		other nodes are comments (size proportional to karma), which are in black unless the
		user comments more than once in the thread.}
	\label{fig:comment_tree}
\end{figure}

\paragraph{Data:} 
{\it Reddit} (\url{https://www.reddit.com}) is a discussion forum with thousands
of sub-communities organized as {\it subreddits}.
Users can initiate a tree-structured discussion thread by making a post in a
subreddit.
Comments are made either directly to the root post or to other comments within
the thread, sometimes triggering sub-discussions.
Each comment can receive upvotes and downvotes from registered users; the
difference is shown as the {\it karma} score beside the comment.
The graph structure of a Reddit disccussion thread
is shown in Fig.\,\ref{fig:comment_tree}.\footnote{Visualization obtained from
	\protect\url{https://whichlight.github.io/reddit-network-vis}.}
In this paper, three popular subreddits are studied: \texttt{AskMen} (1,057K
comments), \texttt{AskWomen} (814K comments), and \texttt{Politics} (2,180K
comments).

\paragraph{Task:}
In many discussion forums, including the those explored here, community endorsement (i.e., karma in Reddit) has a heavy-tailed Zipfian distribution, with most comments getting minimal endorsement and high endorsement comments being rare. Since the high endorsement comments are of most interest, we do not want to treat this as a regression problem using a mean squared error (MSE) objective.\footnote{A prediction error of 50 is minimal for a comment with karma of 500 but substantial for a comment with karma of 1, and the low karma comments dominate the overall MSE.}
Instead, we quantize the karma into $J + 1$ discrete levels and design
a task consisting of $J$ binary classification subtasks which individually
predict whether a comment has karma of at least level-$j$ for each level $j = 1,
\dots, J$ given the text of the comment and the structure of the full discussion thread. (All samples have karma at least level-0.)


\begin{figure}[t]
	\centering
	\includegraphics[trim={1.2cm 1.2cm 0cm 0.5cm},clip,width=0.45\textwidth]{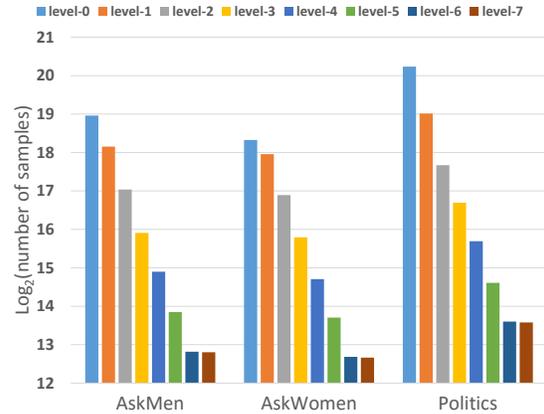}
	\caption{The data distribution for each subreddit.}
	\label{fig:data_dist}
\end{figure}

Karma scores are quantized into 8 levels of community endorsement according to statistics computed over a
large collection of comments in the subreddit.
The quantization process is similar to the head-tail break rule described in
\cite{Jiang2013}.
First, comments with karma no more than 1 are labeled as level-0, indicating
that these comments receive no more upvotes than downvotes.\footnote{The inital
karma score of a comment is 1.}
Then, we compute the median karma score for the rest of the comments, and label
those with below-than-median karma as level-1.  
This process is repeated through level-6, and the remaining comments are labeled
as level-7.
The resulting data distributions are shown in Fig.\,\ref{fig:data_dist}.
Note that the quantization is subreddit dependent, since the distribution and
range of karma tends to vary for different subreddits.

\paragraph{Evaluation metric:} 
Since we use a quantization scheme following a binary thresholding process, we
can compute the F1 score for each level-$j$ subtask ($j = 1, 2, \dots, 7$) by
treating comments whose predicted level is lower than $j$ as negative samples
and others as positive samples.
To evaluate the overall prediction performance, the seven F1 scores are
aggregated via a macro average, which effectively puts a higher weight on the
higher endorsement levels.

\section{Model Description}
\begin{figure}[t]
	\centering
	\includegraphics[width=0.48\textwidth]{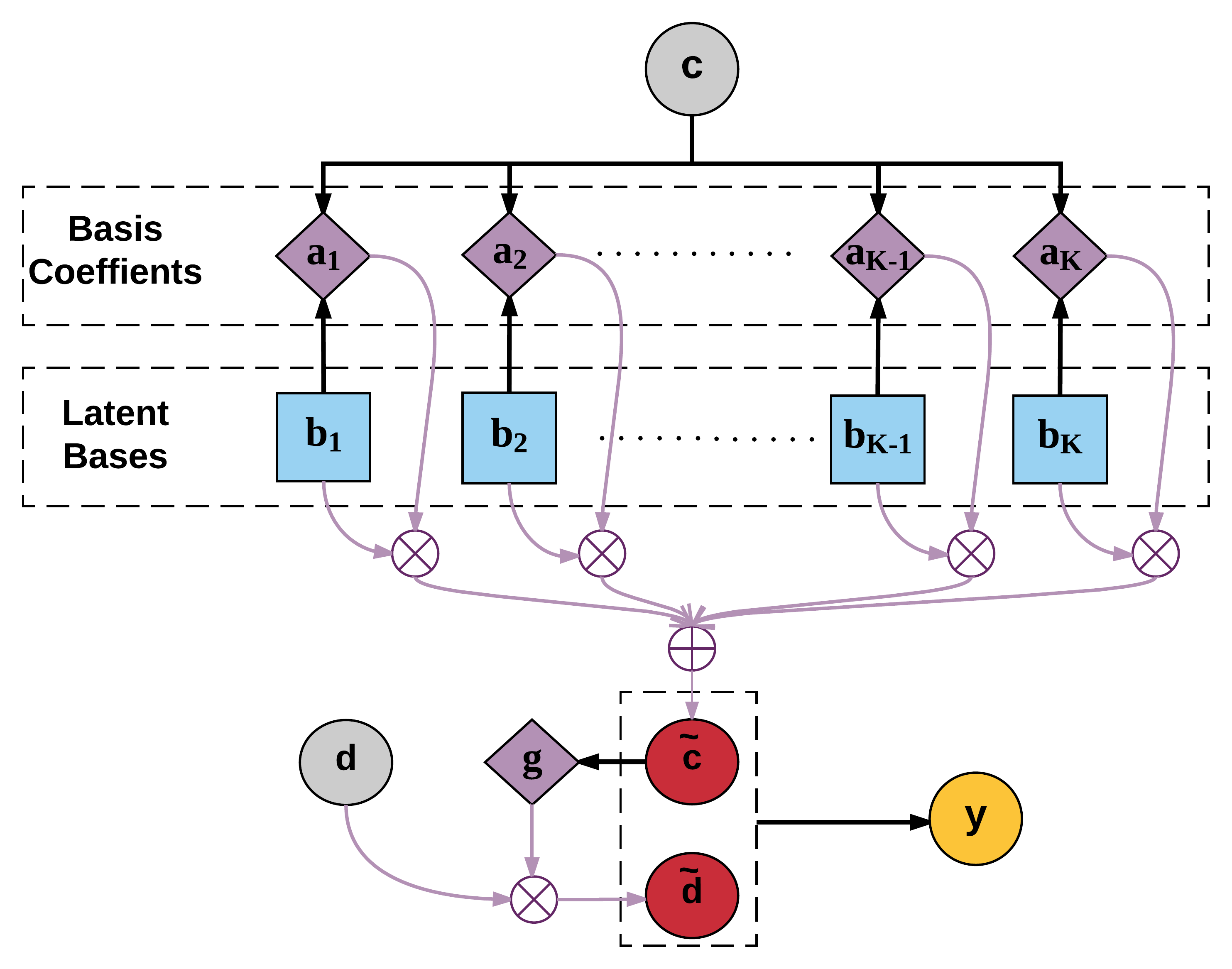}
	\caption{Proposed model:
		Gray circles ${\bf c}$ and ${\bf d}$ are the projected submission
		context features and the encoded textual content vector, respectively.
		Blue boxes ${\bf b}_1, \cdots, {\bf b}_K$ are latent basis vectors, which
		are learned by the neural network.
		Purple diamonds ${\bf a}_1, \cdots, {\bf a}_K$ and ${\bf g}$ represent
		scalers, i.e., the basis coefficients and context-dependent gate value.
		Red circles $\tilde{\bf c}$ and $\tilde{\bf d}$ are the context
		embedding (i.e., a linear combination of latent basis vectors) and the
		weighted text embedding, respectively.  
		The yellow circle ${\bf y}$ is the output layer.
		Black arrows are connections carrying weight matrices.
		$\otimes$ and $\oplus$ indicate multiplication and element-wise addition,
		respectively.
	}
	\label{fig:model}
\end{figure}


The proposed model utilizes two kinds of information for a comment to predict
its quantized karma:~(1)~the submission context encoded by a set of
graph and timing statistics, 
and (2)~the textual content of the comment itself.
Both sources of information are first embedded in a continuous space by a neural
network as illustrated in Fig.\,\ref{fig:model}, where ${\bf c} \in
\mathbb{R}^C$ and ${\bf d} \in \mathbb{R}^D$
encode the submission context and the textual content, respectively. 
As described further below, the two vectors are transformed for use in the final decision function to $\tilde{\bf c}$, a linear combination of latent basis vectors, and $\tilde{\bf d}$, a context-dependent weighted version of the text features.

\begin{table}
	\centering
	{\small
	\begin{tabular}{>{\centering\arraybackslash}m{0.06\textwidth}|m{0.38\textwidth}}
		\hline\hline
		{\bf Range} & {\bf Description} \\
		\hline
		0/1 
		& Whether the comment author is the user who initiated the thread. \\
		\hline
		\multirow{4}{*}[-1.4em]{$\mathbb{Z}_{\geq 0}$} 
		& Number of replies to the comment. \\
		& Number of comments in the subtree rooted from the comment. \\
		& Height of the subtree rooted from the comment. \\
		& Depth of the comment in the tree rooted from the original post. \\
		\hline
		\multirow{2}{*}[-0.5em]{$\mathbb{R}_{\geq 0}$}
		& Relative comment time (in hours) with respect to the original post. \\
		& Relative comment time (in hours) with respect to the parent comment. \\
		\hline\hline
	\end{tabular}}%
	\caption{Features for representing the conversation structure.}
	\label{tab:graph_feats}
\end{table}


\paragraph{Submission context modes:}
Reddit discussions have a variety of conversation structures,
including sections involving many contributors or just a few.
Based on observations that high karma comments seem to co-occur with active
points of discussions, we identify a set of features to represent the submission
context of a comment, specifically aiming to characterize relative timing of the comment within the discussion, participant response to the comment, and whether the comment author is the original poster (see Table\,\ref{tab:graph_feats} for the full list).
The features are normalized to zero mean and unit variance based on the
training set.

In this paper, instead of controlling for the submission context,
we let the model learn latent modes of submission context and examine how the
learned context modes relate to different levels of community endorsement.
The proposed model learns $K$ latent basis vectors ${\bf b}_1,
\cdots, {\bf b}_K \in \mathbb{R}^C$ for characterizing the submission context of
a particular comment in the discussion.
Given the raw submission context feature vector ${\bf x} \in \mathbb{R}^N$,
the model computes a vector ${\bf c} \in \mathbb{R}^C$ as ${\bf c} = {\rm
LReL}({\bf Px})$, where ${\bf P} \in \mathbb{R}^{C \times N}$ is a projection
matrix, and ${\rm LReL}(\cdot)$ is the leaky rectified linear function
\cite{Mass2013ICML} with 0.1 as the slope of the negative part.
Coefficients for these $K$ latent bases are then estimated as 
\begin{align*}
	a_k = {\rm softmax} ( {\bf v}^T {\rm tanh} 
	({\bf U} \left[ {\bf c}; \; {\bf b}_k \right]) ),
\end{align*}
where ${\bf v} \in \mathbb{R}^C$ and ${\bf U} \in \mathbb{R}^{C \times 2C}$ are
parameters to be learned.
The final submission context embedding is obtained as 
$\tilde{\bf c} = \sum_{k=1}^K a_k \cdot {\bf b}_k \in \mathbb{R}^C$.

The computation of basis coefficients is similar to the attention
mechanism that has been used in the context of machine translation
\cite{Bahdanau2015ICLR}, constituency parsing \cite{Vinyals2015NIPS}, question
answering and language modeling \cite{Weston2015ICLR,Sukhbaatar2015NIPS}.
To the best of our knowledge, this is the first attempt to use the attention
mechanism for latent basis learning. 


\paragraph{Text embeddings:}
Recurrent neural networks (RNNs) have been widely used to obtain sequence
embeddings for different applications in recent
years~\cite{Sutskever2014NIPS,Cheng2015EMNLP,Palangi2016TASLP}.
In this paper, we use a bi-directional RNN to encode each sentence, and
concatenate the hidden layers at the last time step of each direction as the
sentence embedding.
For comments with multiple sentences, we average the sentence embeddings into a
single vector as the textual content embedding ${\bf d} \in \mathbb{R}^D$.

For the $t$-th token in a sentence, the hidden layers of the bi-directional RNN
are computed as 
\begin{align*}
	{\bf h}^{(l)}_{t} = {\rm GRU} ({\bf z}_{t}, {\bf h}^{(l)}_{t-1}),
	\quad
	{\bf h}^{(r)}_{t} = {\rm GRU} ({\bf z}_{t}, {\bf h}^{(r)}_{t+1}),
\end{align*}
where ${\bf z}_t \in \mathbb{R}^D$ is the token input vector, ${\bf h}^{(l)}_t \in \mathbb{R}^D$ and ${\bf h}^{(r)}_t \in \mathbb{R}^D$
are the hidden layers for the left-to-right and right-to-left directions,
respectively, 
and ${\rm GRU}(\cdot, \cdot)$ denotes the gated recurrent unit (GRU), which is
proposed by \newcite{Cho2014EMNLP} as a simpler alternative to the long
short-term memory unit \cite{Hochreiter1997} for addressing the vanishing
gradient issue in RNNs.
For consistency of the model and consideration of computation speed, we replace
the hyperbolic tangent function in the GRU with the LReL function.
Although not shown in Fig.\,\ref{fig:model}, weight matrices in the
bi-directional RNN are jointly learned with all other parameters.

To generate the token input vector to the RNN,  we utilize the lemma and
part-of-speech (POS) tag of each token (obtained with the Stanford CoreNLP toolkit \cite{CoreNLP}), in addition to its word form.
A token embedding ${\bf z}_t \in \mathbb{R}^D$ for the $t$-th token in a
sentence is computed as
\begin{align*}
	{\bf z}_t 
	= {\bf E}^{\rm word} {\bf e}_t^{\rm word}
	+ {\bf E}^{\rm pos} {\bf e}_t^{\rm pos}
	+ {\bf E}^{\rm lemma} {\bf e}_t^{\rm lemma},
\end{align*}
where ${\bf e}_t$'s are one-hot encoding vectors for the token, and ${\bf E}$'s
are parameters to be learned. 
The dimensions of these one-hot encoding vectors are determined by the size of the
corresponding vocabularies, which include all observed types except singletons.
Thus, these embedding matrices ${\bf E}$'s have the same first dimension
$D$ but different second dimensions.
This type of additive token embedding has been used in
\cite{Botha2014ICML,Fang2015TASLP} to integrate various types of information
about the token.  
Moreover, it reduces the tuning space since we only need to make a single
decision on the dimensionality of the token embedding.

\paragraph{Gating mechanism:}
For estimating comment karma levels, the textual content should provide
additional information beyond the submission context.
However, we hypothesize that the usefulness of textual content may vary under
different submission contexts since structure reflects size of the readership.
Therefore, we design a context-dependent gating mechanism in the proposed model
to weight the textual factors.
A scalar gate value is estimated from the submission context embedding
$\tilde{\bf c}$, i.e., 
$g = {\rm sigmoid} ({\bf w}^T \tilde{\bf c})$,
where ${\bf w} \in \mathbb{R}^C$ is the parameter to be learned.
The textual content embedding ${\bf d} \in \mathbb{R}^D$ is scaled by the gate
value $g$ before being fed to the output layer, i.e.,
$\tilde{\bf d} = g \cdot {\bf d}$.

\paragraph{Decision function:}
The estimated probability distribution ${\bf y} = [y_0, \dots, y_7]$ over all
quantized karma levels is computed by the softmax output layer, i.e.,
${\bf y} = {\rm softmax} 
( {\bf Q} \left[ \tilde{\bf c}; \; \tilde{\bf d} \right] )$,
where ${\bf Q} \in \mathbb{R}^{J \times (C + D)}$ is the weight matrix to be
learned.
The hypothesized level for a comment is 
$\hat{\mathcal{L}} = {\rm argmax}_j y_j$.
For each level-$j$ subtask, both the label $\mathcal{L}$ and the hypothesis
$\hat{\mathcal{L}}$ are converted to binary values by checking the condition
whether they are no less than $j$.

\section{Parameter Learning}
To train the proposed model, each comment is treated as an independent
sample, and the objective is the maximum log-likelihood of these
samples.
We use mini-batch stochastic gradient descent with a batch size of 32, where the
gradients are computed with the back-propagation algorithm
\cite{Rumelhart1986Nature}.
Specifically, the Adam algorithm is used \cite{Kingma2015ICLR}.
The initial learning rate is selected from the range of [0.0010, 0.0100], with a
step size of 0.0005, according to the log-likelhood of the validation data at
the first epoch.
The learning rate is halved at each epoch once the log-likelihood of the
validation data decreases.
The whole training procedure terminates when the log-likelihood decreases for the
second time.

Each comment is treated as a data sample, and assigned to a partition number in
$\{0, 1, \dots, 9\}$ according to the thread it belongs to.
Each partition has roughly the same number of threads.
We use partitions 4--9 as training data, partitions 2--3 as validation data, and
partitions 0--1 as test data, 
The training data are shuffled at the beginning of each epoch.

As discussed in Section~\ref{sec:data_task}, there are many more low-level
comments than high-level comments, and the evaluation metric effectively puts
more emphasis on high-level comments.
Therefore, rather than using the full training and validation sets, we subsample
the low-level comments (level-0, level-1, level-2, level-3) such that each level
has roughly the same number of samples as level-4.
Since the three subreddits studied in this paper vary in their sizes, to
eliminate the factor of training data size, we use similar amounts of training
($\sim$90K comments) and validation ($\sim$30K comments) data for these
subreddits. 
Note that we do not subsample the test data, i.e., 192K for \texttt{AskMen},
463K for \texttt{AskWomen}, and 1,167K for \texttt{Politics}.

\section{Experiments}
\begin{figure}[t]
	\centering
	\includegraphics[trim={0cm 1cm 0cm 0cm},clip,width=0.49\textwidth]{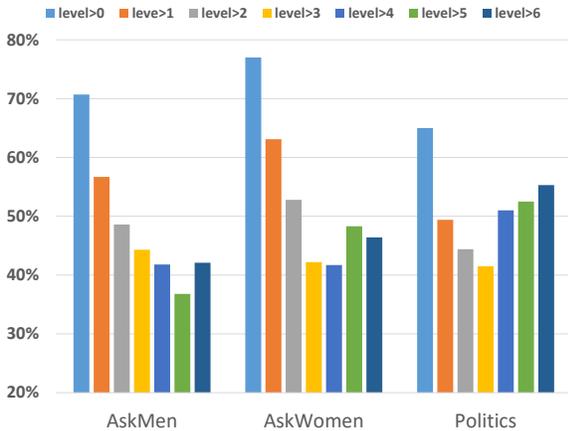}
	\caption{Individual F1 scores for the full model.}
	\label{fig:full_model}
\end{figure}

In this section, we present the performance of the proposed model and conduct
contrastive experiments to
study model variants in two dimensions. 
For the submission context features, we compare 
representations obtained via feedforward neural networks to that obtained by a
learned combination of latent basis vectors. In terms of textual features, we compare a model which uses no text, context-independent text features, and a context-depending gating mechanism.
Finally, we analyze the learned latent submission context modes,
as well as context-dependent gate values that reflect the amount of textual
information used by the full model.

\subsection{Model Configuration}
All parameters in the neural networks except bias terms are initialized
randomly according to the Gaussian distribution $\mathcal{N}(0, 10^{-2})$.
We tune the number of latent bases $K$ and the number of hidden layer neurons
$C$ and $D$ based on the macro F1 scores on the validation data. 
For the full model, the best configuration uses 
$K = 8$, $C = 32$ and $D = 64$ for all subreddits, except \texttt{Politics}
where $D = 32$.

\subsection{Main Results}


\begin{table}[t]
	\centering
	{\small
	\begin{tabular}{l|ccc}
		\hline\hline
		& \texttt{AskMen}
		& \texttt{AskWomen}
		& \texttt{Politics} \\
		\hline
		SubtreeSize 		& 39.1 & 42.9 & 41.7 \\
		ConvStruct 			& 43.9 & 41.4 & 42.0 \\
		\hline
		Feedfwd-1 			& 46.5 & 50.6 & 49.6 \\
		Feedfwd-2 			& 46.8 & 50.9 & 49.8 \\
		Feedfwd-3 			& 47.1 & 50.5 & 50.0 \\
		LatentModes 		& 47.0 & 51.0 & 50.3 \\
		\hline\hline
	\end{tabular}}%
	\caption{Test macro F1 scores for models that do not use the textual content
		information.}
	\label{tab:exp_attn}
\end{table}

\begin{table}[t]
	\centering
	{\small
	\begin{tabular}{c|ccc}
		\hline\hline
		& \texttt{AskMen}
		& \texttt{AskWomen}
		& \texttt{Politics} \\
		\hline
		No text		& 47.0 & 51.0 & 50.3 \\
		Un-gated 	& 48.3 & 52.5 & 49.5 \\
		Gated  		& {\bf 48.7} & {\bf 53.1} & {\bf 51.3} \\
		\hline\hline
	\end{tabular}}%
	\caption{Test macro F1 scores for models with and without the gating
	mechanism. 
	All models use latent modes to represent the submission context information.}
	\label{tab:exp_gate}
\end{table}

\begin{figure*}[t]
	\centering
	\subfloat{
		\includegraphics[trim={2.5cm 10.5cm 0cm 10cm},clip,width=0.7\textwidth]
		{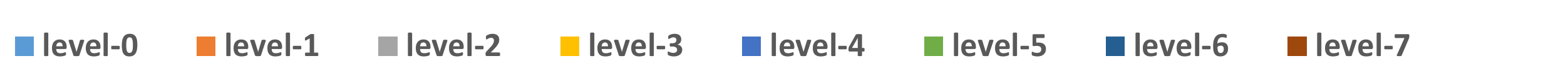}
	}
	\\
	\addtocounter{subfigure}{-1}
	\subfloat[\texttt{AskMen}]{
		\includegraphics[trim={0.5cm 1cm 1cm 1cm},clip,width=0.32\textwidth]
		{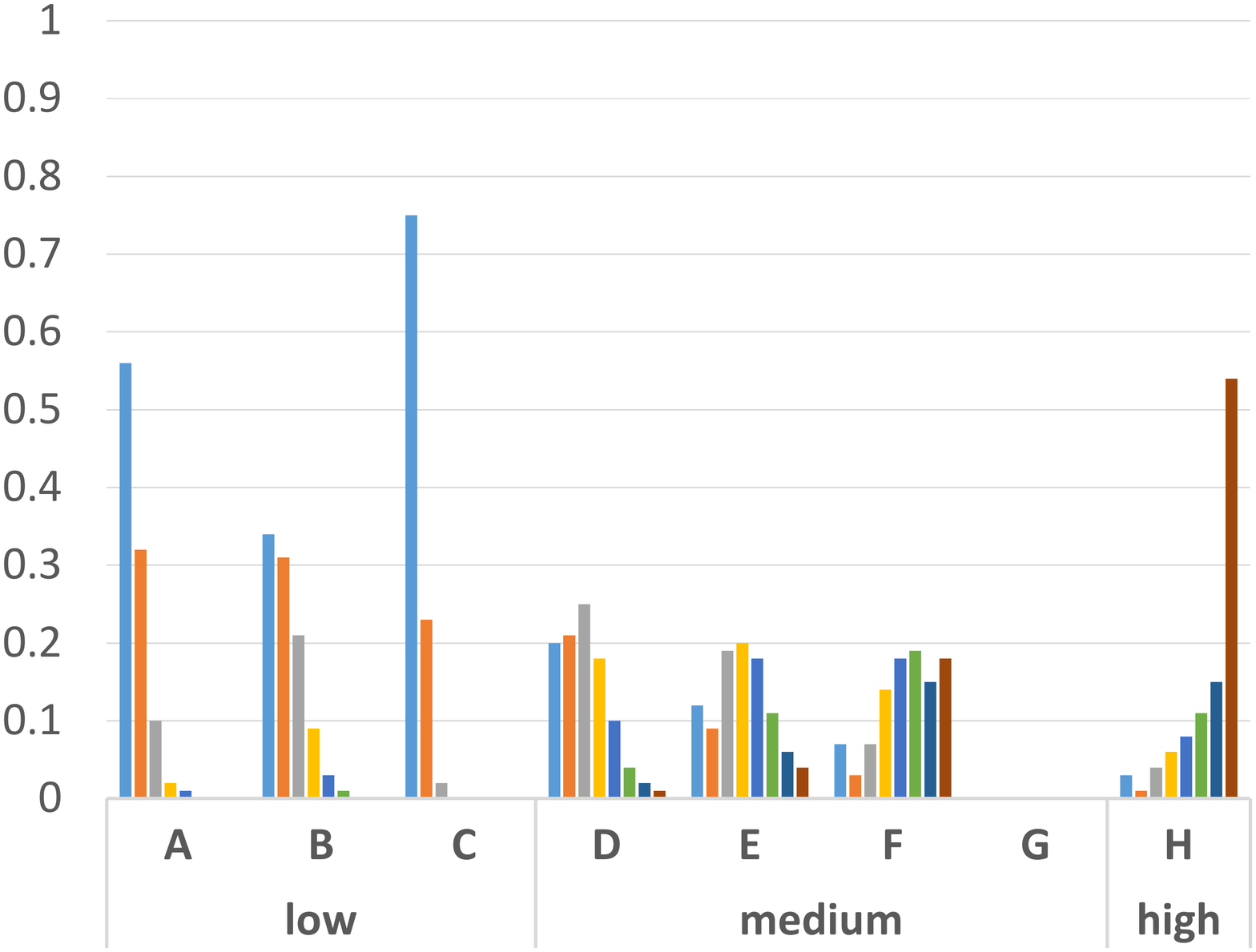}
	}%
	\subfloat[\texttt{AskWomen}]{
		\includegraphics[trim={0.5cm 1cm 1cm 1cm},clip,width=0.32\textwidth]
		{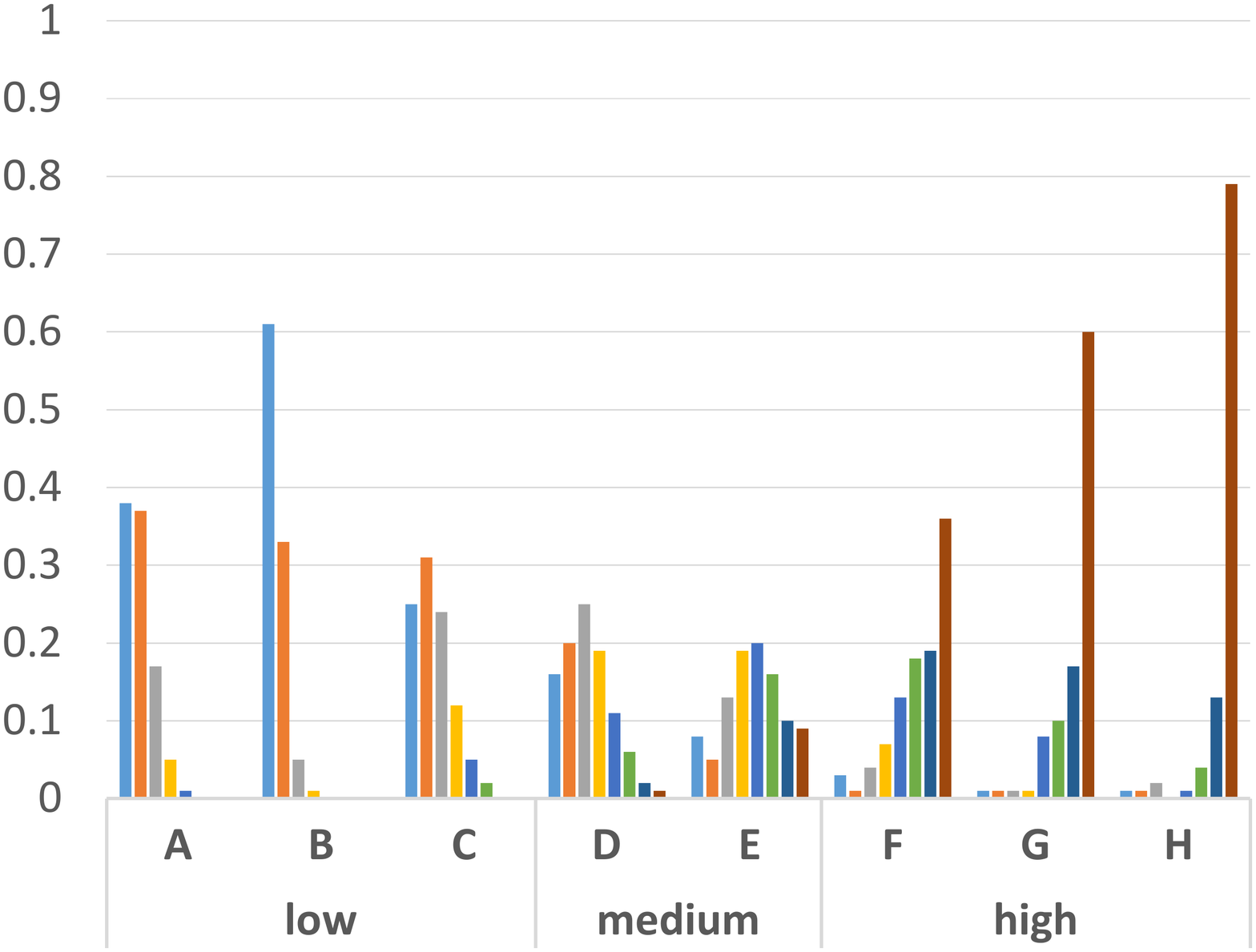}
	}%
	\subfloat[\texttt{Politics}]{
		\includegraphics[trim={0.5cm 1cm 1cm 1cm},clip,width=0.32\textwidth]
		{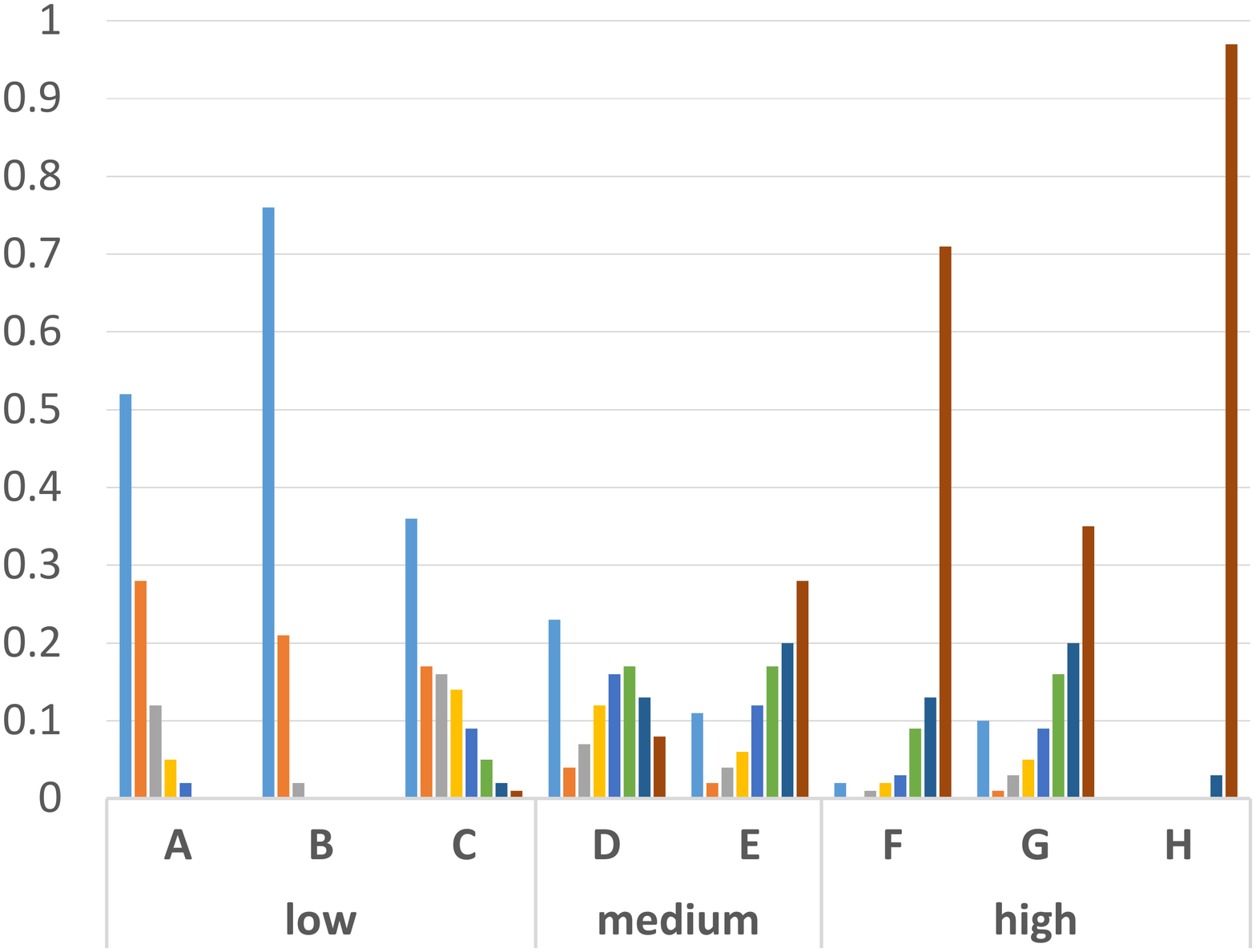}
	}
	\caption{Empirical distributions of levels for each latent mode.
	Modes are grouped by dominating levels, i.e., level-0 and level-1 as
	\texttt{low}, level-6 and level-7 as \texttt{high}, and the rest as
	\texttt{medium}.
	Within each cluster, the modes are sorted by the number of samples.}
	\label{fig:cluster_distrib}
\end{figure*}
\begin{figure*}[t]
	\centering
	\subfloat[\texttt{AskMen}]{
		\includegraphics[trim={1.5cm 1.0cm 2.0cm 1.0cm},clip,width=0.32\textwidth]
		{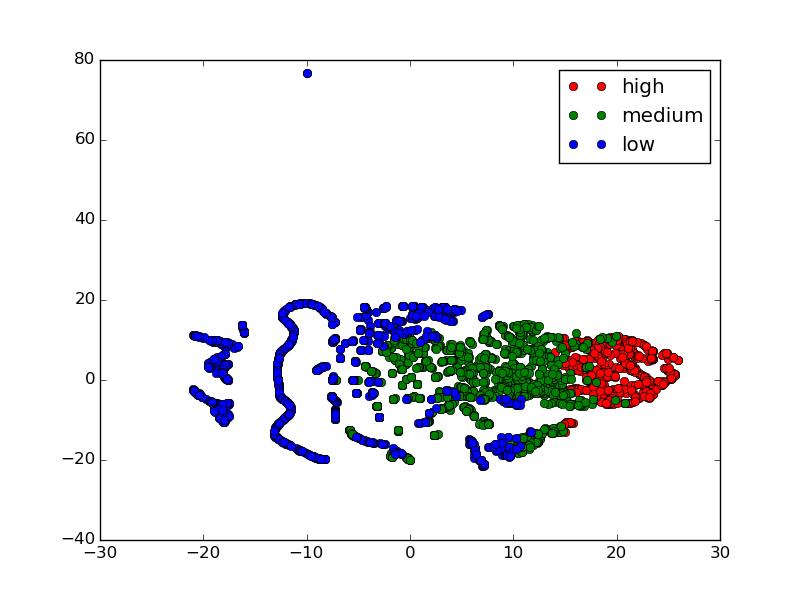}
	}%
	\subfloat[\texttt{AskWomen}]{
		\includegraphics[trim={1.5cm 1.0cm 2.0cm 1.0cm},clip,width=0.32\textwidth]
		{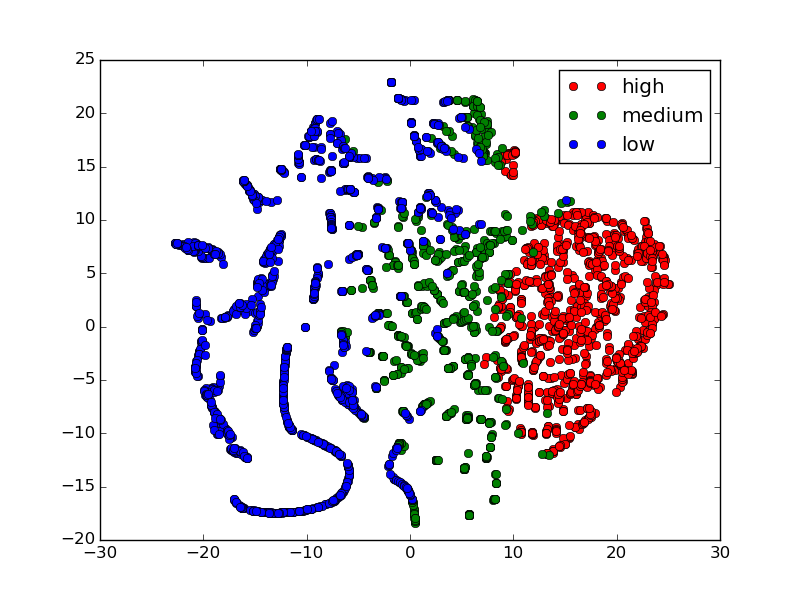}
	}%
	\subfloat[\texttt{Politics}]{
		\includegraphics[trim={1.5cm 1.0cm 2.0cm 1.0cm},clip,width=0.32\textwidth]
		{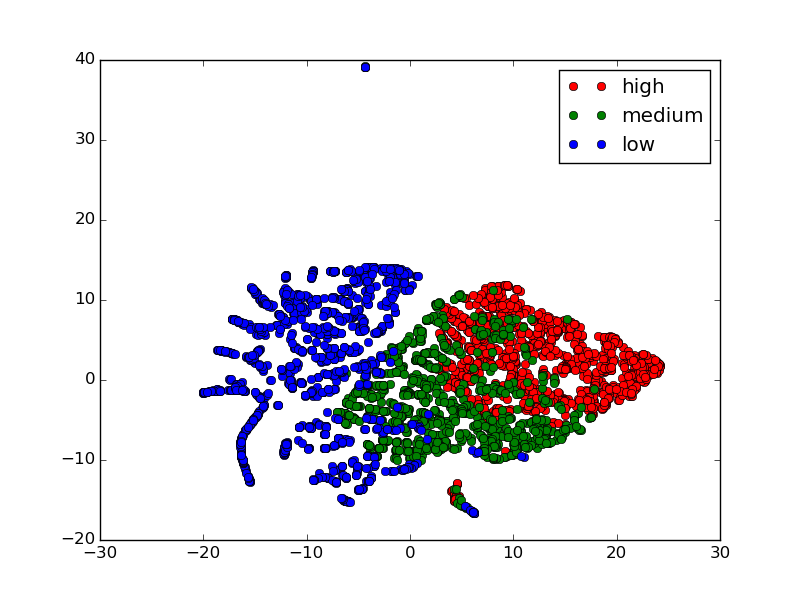}
	}%
	\caption{Visualization of learned clusters.}
	\label{fig:cluster_tsne}
\end{figure*}

\begin{figure}[t]
	\centering
	\includegraphics[trim={0.8cm 1.0cm 0.5cm 1.0cm},clip,width=0.49\textwidth]{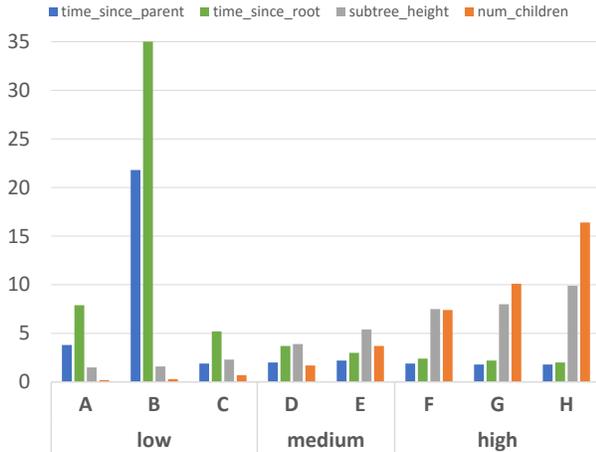}
	\caption{Mean values of four submission context features for each latent mode
		of \texttt{AskWomen}.}
	\label{fig:cluster_stat_askwomen}
\end{figure}

The performance of the full model on individual levels is presented in
Fig.\,\ref{fig:full_model}.
As expected, the lowest level comments are easier to classify.
Detection of high-level comments is most reliable in the \texttt{Politics} subreddit, but still difficult.

Table\,\ref{tab:exp_attn} compares models variants that only use the submission
context features.
The \texttt{SubtreeSize} baseline uses a multinominal logistic regression model
to predict the level according to the subtree size feature alone, whereas the
\texttt{ConvStruct} uses the same model but with all conversation structure features
defined in Tabel\,\ref{tab:graph_feats}. All baselines are stronger than predicting based on prior distributions, which has F1 scores in the 11-17 range.
The model \texttt{Feedfwd-n} is a feedforward neural network with \texttt{n}
hidden layers; it uses the submission context feature ${\bf c}$ in
Fig.\,\ref{fig:model} for prediction.
The model \texttt{LatentBases} represents the submission context information by
a linear combination of latent bases; it uses $\tilde{\bf c}$ in
Fig.\,\ref{fig:model} for prediction.
Compared with \texttt{Feedfwd-1} in terms of the number of model parameters,
\texttt{Feedfwd-2}, \texttt{Feedfwd-3} and \texttt{LatentBases} have $C^2$, 
$2C^2$, and $(2C^2 + K)$ extra parameters, respectively.
These models have similar performance, but
there is a slight improvement by increasing model capacity.
While the proposed method does not give a significant performance gain, it leads
to a more interpretable model. 

Table\,\ref{tab:exp_gate} studies the effect of adding text and introducing the gating
mechanism.
The un-gated variant uses ${\bf d}$ instead of $\tilde{\bf d}$ 
for prediction.
Without the gating mechanism, textual information provides significant
improvement for \texttt{AskMen} and \texttt{AskWomen} but not for \texttt{Politics}. 
With the introduced dynamic gating mechanism, the textual
information is used more effectively for all three subreddits.

\subsection{Analysis}
In this subsection, we analyze the learned submission context modes and the gate values that control the amount of textual information
to be used by the model for predicting comment karma level.

\paragraph{Submission context modes:}
To study the submission context modes, we assign each comment to a cluster
according to which basis vector receives the highest weight: ${\rm argmax}_{k=1, \dots, K} a_k$.
The label distribution for each cluster is shown in
Fig.\,\ref{fig:cluster_distrib}.
It can be observed that some clusters are dominated by level-0 comments, and
others are dominated by level-7 comments.
In Fig.\,\ref{fig:cluster_tsne}, we visualize the learned clusters
by projecting the raw conversation structure features ${\bf x}$ to a 2-dimensional space
using the t-SNE algorithm \cite{Maaten2008MLR}.
For purposes of illustrating cross-domain similarities, we group the clusters
dominated by level-0 and level-1 comments into a low endorsement cluster, those
dominated by level-6 and level-7 into a high endorsement cluster, and the rest
as the medium endorsement cluster.
It can be seen that the learned clusters split the comments with a
consistent pattern, with the higher endorsement comments towards the right and the low endorsement comments to the left.


In Fig.\,\ref{fig:cluster_stat_askwomen}, we show mean values of four selected
submission context features for each latent mode of \texttt{AskWomen}, where units of time are in hours.
High karma comments tend to be submitted early in the discussion, and the number of children (direct replies) is similar to or greater than the height of its subtree (corresponding to a broad subtree). Low and medium karma comments have a ratio of number of children to subtree height that is less than one. Low karma comments tend to come later in the discussion overall (time since root) but also later in terms of the group of responses to a parent comment (time since parent). These trends hold for all three subreddits. All subreddits have within-group differences in the mode characteristics, particularly the low-karma modes. For \texttt{AskWomen}, graph cluster B corresponds to comments made at the end of a discussion, which are more likely to be low karma because there are fewer readers and less opportunity for a new contribution. Cluster C comments come earlier in the discussion but have small subtrees compared to other early comments.




\begin{table}[t]
	\centering
	\begin{tabular}{c|c|c|c}
		\hline\hline
		& \texttt{AskMen} & \texttt{AskWomen} & \texttt{Politics} \\
		\hline
		medium & 0.87 & 0.87 & 0.85 \\
		high & 0.67 & 0.66 & 0.76 \\
		\hline\hline
	\end{tabular}
	\caption{Text gate values relative to low karma modes.}
	\label{tab:relative_text_gate}
\end{table}

\paragraph{Text gate:} 
In Table\,\ref{tab:relative_text_gate}, we show the mean gate values $g$ for each group of
latent modes. Since gate values are not comparable across subreddits due to dynamic range of feature values, the values shown are scaled by the value for the low-level mode.
We observe a consistent trend across all
subreddits: lower gate values for higher karma.
Recall that the high karma comments typically spawn active discussions. 
Thus, a possible explanation is that users may be biased to endorse comments that others are endorsing, 
making the details of the content less important.

\section{Conclusion}
In summary, this work has addressed the problem of  predicting community
endorsement of comments in a discussion forum using a new neural network
architecture that integrates submission context features (including relative
timing and response structure) with features extracted from the text of a
comment. The approach represents the submission context in terms of a linear
combination of latent basis vectors that characterize the dynamic conversation
mode, which gives results similar to using a deep network but is more
interpretable. The model also includes a dynamic gate for the text content,
and analysis shows that when response structure is available to the predictor,
the content of a comment has the most utility for comments that are not in
active regions of the discussion. These results are based on characterizing quantized levels of karma with a series of binary classifiers. Quantized karma prediction could also be framed as an ordinal regression task, which would involve a straightforward change to the neural network learning objective.

This work differs from related work on popularity prediction in that the task
does not control for content of a post/comment, nor limit the time window of the
submission. With fewer controls, it is more difficult to uncover the aspects of
textual content that contribute to endorsement, but by conditioning on
submission context we can begin to understand herd effects of endorsement.  The
task described here also differs from previous work in that the full (or almost
full) discussion thread is available in extracting features characterizing the
response to the comment, but the modeling framework would also be useful with a
limited window lookahead. The results using the full discussion tree also show
the limits of using response volume to measure endorsement.

A limitation of this work is that the submission context is represented only in
terms of the relative timing and graph structure in a discussion thread and does
not use the text within earlier or responding comments. Prior work has shown
that the relevance of a comment to the preceding discussion matters
\cite{Jaech2015EMNLP}, and clearly the sentiment expressed in responses should
provide important cues. Capturing these different sources of information in a
gated framework is of interest for future work.

\subsection*{Acknowledgments}
This paper is based on work supported by the DARPA DEFT Program. Views expressed are those of the authors and do not reflect the official policy or position of the Department of Defense or the U.S.\ Government. 


\end{document}